\pdfoutput=1
\documentclass[conference]{IEEEtran}
\usepackage{mathtools} 
\usepackage{amsthm, amssymb} 
\usepackage{bm} 
\usepackage{microtype} 
\usepackage{siunitx} 
\usepackage[noadjust]{cite}
\usepackage[hidelinks]{hyperref} 
\usepackage[nameinlink]{cleveref} 
\usepackage{orcidlink} 

\hypersetup{hidelinks, pdfinfo = {Title = {Coded Caching Does Not Generally Benefit From Selfish Caching}, Author = {Federico Brunero and Petros Elia}, Subject = {Information-Theoretic Caching}, Keywords = {Coded Caching, File Popularity, Index Coding, Information-Theoretic Converse, Selfish Caching.}}}
\allowdisplaybreaks

\Crefformat{figure}{#2Fig.~#1#3}

\theoremstyle{plain}
\newtheorem{theorem}{Theorem}
\newtheorem{lemma}{Lemma}
\newtheorem{corollary}{Corollary}[theorem]

\theoremstyle{definition}
\newtheorem{definition}{Definition}
\newtheorem{example}{Example}

\IEEEoverridecommandlockouts

\title{Coded Caching Does Not Generally Benefit From Selfish Caching\thanks{This work was supported by the European Research Council (ERC) through the EU Horizon 2020 Research and Innovation Program under Grant 725929 (Project DUALITY). An extended version of this work~\cite{Brunero2021UnselfishCodedCaching} has been submitted to IEEE Transactions on Information Theory.}}

\author{\IEEEauthorblockN{Federico Brunero\textsuperscript{\orcidlink{0000-0002-6980-3827}} and Petros Elia\textsuperscript{\orcidlink{0000-0002-3531-120X}}}
\IEEEauthorblockA{Communication Systems Department, EURECOM, Sophia Antipolis, France\\
Email: \{brunero, elia\}@eurecom.fr}}

\begin{document}

\bstctlcite{IEEEexample:BSTcontrol}

\nocite{MaddahAli2014FundamentalLimitsCaching, Parrinello2020FundamentalLimitsCoded, Lampiris2021ResolvingFeedbackBottleneck, Toelli2020MultiAntennaInterference, Zhang2018CodedCachingArbitrary, Zhang2021DeepLearningWireless, Hachem2017CodedCachingMulti, Lampiris2020FullCodedCaching, Joudeh2021FundamentalLimitsWireless}

\maketitle

\begin{abstract}

In typical coded caching scenarios, the content of a central library is assumed to be of interest to all receiving users. However, in a realistic scenario the users may have diverging interests which may intersect to various degrees. What happens for example if each file is of potential interest to, say, $\SI{40}{\percent}$ of the users and each user has potential interest in $\SI{40}{\percent}$ of the library? What if then each user caches selfishly only from content of potential interest? In this work, we formulate the symmetric selfish coded caching problem, where each user naturally makes requests from a subset of the library, which defines its own file demand set (FDS), and caches selfishly only contents from its own FDS. For the scenario where the different FDSs symmetrically overlap to some extent, we propose a novel information-theoretic converse that reveals, for such general setting of symmetric FDS structures, that selfish coded caching yields a load performance which is strictly worse than that in standard coded caching.

\end{abstract}


\section{Introduction}

The explosion of network traffic in the recent years has sparked much interest in new communication techniques with the purpose of reducing the traffic load. In this context, caching has always played a key role in bringing contents closer to their destinations, thus reducing the volume of the communication problem during peak hours. Coded caching~\cite{MaddahAli2014FundamentalLimitsCaching} has been proposed as a clever way to better exploit caching capabilities of receiving users, where such coded technique represents, in comparison to traditional prefetching, a key breakthrough in the way end-user caches are employed to change both the \emph{volume} of the communication problem and the \emph{structure} of the problem itself. Current research on coded caching encompasses many topics such as the role of multiple antennas in caching~\cite{Parrinello2020FundamentalLimitsCoded, Lampiris2021ResolvingFeedbackBottleneck, Toelli2020MultiAntennaInterference}, the interplay between caching and file popularity~\cite{Zhang2018CodedCachingArbitrary, Zhang2021DeepLearningWireless, Hachem2017CodedCachingMulti} and a variety of other scenarios~\cite{Lampiris2020FullCodedCaching, Joudeh2021FundamentalLimitsWireless}. For a thorough review of the existing coded caching works, we strongly encourage the reader to refer to the longer version of this work~\cite{Brunero2021UnselfishCodedCaching}.

\subsection{Heterogeneous User Profiles and Selfish Coded Caching}

If on one hand a key ingredient in standard prefetching systems has commonly been the exploitation of the fact that some contents are more popular than others, and thus are generally to be allocated more cache space, on the other we are just beginning to explore the interplay between coded caching, heterogeneous preferences and selfish caching, where by selfish caching we refer to caching policies where each user caches only contents from its own set of individual preferences.

Recent works have sought to explore this interplay. For example, in the context of coded caching with users having heterogeneous content preferences, the work in~\cite{Wang2019CodedCachingHeterogeneous} analyzed the peak load of three different coded caching schemes that account for the user preferences, revealing occasional performance gains that are similarly dependent on the structure of these preferences. Related analysis appears in~\cite{Zhang2020AverageRateCoded}, now for the average load of the same schemes in~\cite{Wang2019CodedCachingHeterogeneous}. On the other hand, the work in~\cite{Chang2019CodedCachingHeterogeneous} focused on finding instances where unselfish coded caching outperforms selfish designs. This work nicely considered the performance of selfish coded caching in the context of heterogeneous file demand sets (FDSs), cleverly employing bounds to show that, for the case of $K = 2$ users and $N = 3$ files, unselfish designs strictly outperform selfish designs in terms of communication load, albeit only by a factor of up to $\SI{14}{\percent}$. In addition, the notable work in~\cite{Chang2020CodedCachingTwo} established, under the assumption of selfish and uncoded prefetching, the optimal average load for the case of $K = 2$ users and a variety of overlaps between the two users' profiles, also providing explicit prefetching schemes of a selfish nature. In a similar vein, the recent work in~\cite{Wan2021OptimalLoadMemory} considered the scenario where users are interested in a limited set of contents depending on their location. To the best of our knowledge, the above constitutes the majority of works on selfish coded caching.

\subsection{Main Contributions}

First, we propose a novel symmetric FDS structure which aims to regulate the selfishness effect and to encapsulate the aspect of overlapping of interests by calibrating the degree of separation between the interests of the users.

Secondly, we develop for the formulated system model an information-theoretic converse on the optimal worst-case load under the assumption of uncoded and selfish placement. The interesting outcome of such bound is that coded caching does not generally benefit from selfish caching policies. In~\cite{Brunero2021UnselfishCodedCaching} we also provide instances where this bound is found to be tight.

\subsection{Paper Outline}

The system model is presented in~\Cref{sec: System Model}. \Cref{sec: Main Results} presents the information-theoretic converse, whose main proof in~\Cref{sec: Main Proof} is followed by the clarifying example in~\Cref{sec: A Detailed Example for the Converse Bound}. \Cref{sec: Conclusions} concludes the paper.

\subsection{Notation}\label{sec: Notation}  

We denote by $\mathbb{Z}^{+}$ the set of positive integers. For $n \in \mathbb{Z}^{+}$, we define $[n] \coloneqq \{1, 2, \dots, n\}$. If $a, b \in \mathbb{Z}^{+}$ such that $a < b$, then $[a : b] \coloneqq \{a, a + 1, \dots, b - 1, b\}$. For sets we use calligraphic symbols, whereas for vectors we use bold symbols. Given a finite set $\mathcal{A}$, we denote by $|\mathcal{A}|$ its cardinality. We denote by $\binom{n}{k}$ the binomial coefficient and we let $\binom{n}{k} = 0$ whenever $n < 0$, $k < 0$ or $n < k$. We use $m \bmod n$ to denote the modulo operation on $m$ with integer divisor $n$, letting $m \bmod n = n$ when $n$ divides $m$. For $n \in \mathbb{Z}^{+}$, we denote by $S_n$ the group of all permutations of $[n]$ and by $H_n$ the group of circular\footnote{This means that no vector $(\pi(1), \dots, \pi(n))$ with $\pi \in H_n$ can be obtained as a rotation of $(\sigma(1), \dots, \sigma(n))$ for some $\sigma \in H_n$ with $\sigma \neq \pi$.} permutations of $[n]$. To simplify, we denote by $\bm{\pi}^n \coloneqq (\pi(1), \dots, \pi(n))$ the vector of elements from $[n]$ permuted according to $\pi \in S_n$ and we let $\pi^{-1}$ denote the inverse function of $\pi$.

\section{System Model}\label{sec: System Model}

We consider the centralized caching scenario in~\cite{MaddahAli2014FundamentalLimitsCaching} where one central server has access to a library $\mathcal{L}$ containing $N$ files of $B$ bits each. This server is connected to $K$ users through a shared error-free broadcast channel, and each user is equipped with a cache of size $M$ files or, equivalently, $MB$ bits.

The system works in two phases, i.e., the placement phase and the delivery phase. During the placement phase, the end-user caches are filled by the server according to a caching policy without any knowledge of future requests. During the delivery phase, after the demands of the users are simultaneously revealed, the server sends coded messages over the shared link to deliver the missing information to each user. Assuming that in the delivery phase each user demands simultaneously one file, the worst-case communication load $R$ is defined as the total number of transmitted bits, normalized by the file-size $B$, that can guarantee delivery of all requested files in the worst-case scenario. The optimal worst-case communication load $R^{\star}$ is then formally defined as
\begin{equation}
  R^{\star}(M) \coloneqq \inf \{R : (M, R) \textnormal{ is achievable}\}
\end{equation}
where the tuple $(M, R)$ is said to be \emph{achievable} if there exists a caching-and-delivery scheme which guarantees, for any possible demand, a load $R$ for a given memory value $M$.

To capture the interplay between coded caching, heterogeneous interests and selfish caching, we propose an FDS structure that allows to calibrate the degree of separation between the interests of the different users.

\begin{definition}[The Symmetric $(K, \alpha, F)$ FDS Structure]
 Let $\alpha \in [K]$ and $F \in \mathbb{Z}^{+}$. The symmetric $(K, \alpha, F)$ FDS structure assumes an $N$-file library $\mathcal{L} = \{\mathcal{W}_{\mathcal{S}} : \mathcal{S} \subseteq [K], |\mathcal{S}| = \alpha \}$ to be a collection of disjoint file classes 
 $\mathcal{W}_{\mathcal{S}} = \{ W_{f, \mathcal{S}} : f \in [F]\}$, where each class $\mathcal{W}_{\mathcal{S}}$ consists of $F$ different files and each of such files is of interest to the $\alpha$ users in the set $\mathcal{S}$. As a consequence, the FDS of each user $k \in [K]$, which describes the files this user is potentially interested in, is defined as
  \begin{equation}
    \mathcal{F}_{k}  = \left\{\mathcal{W}_{\mathcal{S}} : \mathcal{S} \subseteq [K], |\mathcal{S}| = \alpha, k \in \mathcal{S} \right\}.
  \end{equation}
\end{definition}

We can see from the above that the library is partitioned into $C = \binom{K}{\alpha}$ disjoint classes of files, each composed of $F$ files, for a total of $N = FC$ files. Each user $k \in [K]$ is then interested in its own FDS, which is made of $F\binom{K - 1}{\alpha - 1}$ files for each $k \in [K]$. Each file class is identified by an $\alpha$-tuple $\mathcal{S}$, where $\mathcal{S}$ simply tells which $\alpha$ users are interested in the file class $\mathcal{W}_{\mathcal{S}}$. We focus on the regime $N \geq K$, which is equivalent to $F \geq \lceil K/C \rceil$ for any fixed $K$ and $\alpha$. The following example can help familiarize the reader with the notation.
 
 \begin{example}[The Symmetric $(4, 3, 2)$ FDS Structure]
  Consider the symmetric $(K, \alpha, F) = (4, 3, 2)$ FDS structure. There are $C = \binom{K}{\alpha} = 4$ classes\footnote{For simplicity, we will omit braces and commas when indicating sets, such that for example $\mathcal{W}_{\{1, 2, 3\}}$ may be written as $\mathcal{W}_{123}$.} $\mathcal{W}_{123}, \mathcal{W}_{124}, \mathcal{W}_{134}, \mathcal{W}_{234}$ and $N = 8$ files: $W_{1, 123}$ and $W_{2, 123}$ from class $\mathcal{W}_{123}$, then $W_{1, 124}$ and $W_{2, 124}$ from class $\mathcal{W}_{124}$, and so on. The $K$ FDSs take the form
  \begin{alignat}{2}
    \mathcal{F}_{1} &= \{\mathcal{W}_{123}, \mathcal{W}_{124}, \mathcal{W}_{134}\} & \ \mathcal{F}_{3} &= \{\mathcal{W}_{123}, \mathcal{W}_{134}, \mathcal{W}_{234}\} \\
    \mathcal{F}_{2} &= \{\mathcal{W}_{123}, \mathcal{W}_{124}, \mathcal{W}_{234}\} & \ \mathcal{F}_{4} &= \{\mathcal{W}_{124}, \mathcal{W}_{134}, \mathcal{W}_{234}\}
  \end{alignat}
  where each FDS consists of $6$ files in total. For $\mathcal{F}_k$ denoting the FDS of user $k \in [4]$, user~$1$ is interested in files in $\mathcal{F}_1 = \{W_{f, \mathcal{S}}: \mathcal{S} \subseteq [4], |\mathcal{S}| = 3, 1 \in \mathcal{S}, f \in [2]\}$, user~$2$ is interested in files in $\mathcal{F}_2 = \{W_{f, \mathcal{S}}: \mathcal{S} \subseteq [4], |\mathcal{S}| = 3, 2 \in \mathcal{S}, f \in [2]\}$, and so on.
\end{example}

Deviating from standard notation practices, we will use the double-index notation $W_{f_k, \mathcal{D}_k}$ to denote the file requested by user $k$. Hence, any demand will be described by the tuple $(\bm{d}, \bm{f})$ with $\bm{d} = (\mathcal{D}_1, \dots, \mathcal{D}_K)$ and $\bm{f} = (f_1, \dots, f_K)$. Trivially, it holds that $W_{f_k, \mathcal{D}_k} \in \mathcal{F}_k$ for each $k \in [K]$.

Our goal is to provide a converse bound on the optimal worst-case load under the assumption of uncoded and \emph{selfish} placement, whose definition is given in the following.

\begin{definition}[Uncoded and Selfish Cache Placement]\label{def: Uncoded and Selfish Cache Placement}
A cache placement is \emph{uncoded} if the bits of the files are simply copied within the caches of the users and is \emph{selfish} when it guarantees for each $k \in [K]$ that a subfile of $W_{f, \mathcal{S}}$ can be cached at user $k$ only if $W_{f, \mathcal{S}} \in \mathcal{F}_k$, i.e., only if the file $W_{f, \mathcal{S}}$ is potentially of interest to user $k$.
\end{definition}

We denote by $R^\star_{\textnormal{u}, \textnormal{s}}$ the optimal worst-case load under uncoded and selfish placement, and by $R_{\textnormal{MAN}}$ the load of the Maddah-Ali and Niesen (MAN) scheme in~\cite{MaddahAli2014FundamentalLimitsCaching}.

\section{Main Results}\label{sec: Main Results}

The converse bound employs the index coding techniques of~\cite{Wan2020IndexCodingApproach} that proved the optimality of the MAN scheme under the constraint of uncoded cache placement. Our main challenge will be to adapt the index coding approach to reflect the symmetric $(K, \alpha, F)$ FDS structure proposed in~\Cref{sec: System Model}. The result is stated in the following theorem, whereas the proof is presented in~\Cref{sec: Main Proof}.

\begin{theorem}\label{thm: Converse Bound for Coded Caching under Uncoded and Selfish Prefetching}
  Under the assumption of uncoded and selfish cache placement, and given the symmetric $(K, \alpha, F)$ FDS structure, the optimal worst-case communication load $R^{\star}_{\textnormal{u}, \textnormal{s}}$ is lower bounded by $R_{\textnormal{LB}}$ which is a piecewise linear curve with corner points
  \begin{equation}\label{eqn: Information-Theoretic Lower Bound}
    (M, R_{\textnormal{LB}}) = \left(t\frac{N}{K}, \frac{\binom{\alpha}{t + 1} + (K - \alpha)\binom{\alpha - 1}{t}}{\binom{\alpha}{t}}\right), \ \forall t \in [0 : \alpha]
  \end{equation}
corresponding to 
  \begin{equation}
    R_{\textnormal{LB}} = \frac{K(1 - \gamma_\alpha)}{K\gamma + 1}\Big[(K - \alpha)\gamma + 1\Big]
  \end{equation}
  where $\gamma \coloneqq M/N$ and $\gamma_\alpha \coloneqq K \gamma / \alpha$.
\end{theorem}

The converse bound presented here shows that adding the selfish cache placement constraint implies a higher optimal communication load compared to the unselfish scenario. This is nicely reflected in the following corollary.

\begin{corollary}\label{cor: Comparison between Converse Bound and MAN Scheme}
Given the symmetric $(K, \alpha, F)$ FDS structure and $\alpha \in [2 : K - 1]$, the converse reveals that 
\begin{align}
    \frac{R^{\star}_{\textnormal{u}, \textnormal{s}}(t)}{R_{\textnormal{MAN}}(t)} & \geq 1, \quad \forall t \in [0 : \alpha - 1] \\
    \frac{R^{\star}_{\textnormal{u}, \textnormal{s}}(t)}{R_{\textnormal{MAN}}(t)} & > 1, \quad \forall t \in [\alpha - 2].
\end{align}
In the non-trivial range $t \in [0 : \alpha - 1]$, selfish coded caching is not better than unselfish coded caching, while optimal unselfish coded caching, in the non-extremal points of $t$ and under uncoded placement, strictly outperforms any implementation of selfish coded caching.
\end{corollary}
\begin{proof}
    Due to lack of space, the proof is relegated to the longer version of this work~\cite{Brunero2021UnselfishCodedCaching}.
\end{proof}

In the above comparison we excluded $\alpha \in \{1, K\}$. Indeed, when $\alpha = 1$ the comparison is trivial, whereas for $\alpha = K$ the converse expression naturally matches that of unselfish coded caching.

\subsection{Main Proof}\label{sec: Main Proof}

The derivation of the converse makes extensive use of the connection between caching and index coding identified in~\cite{MaddahAli2014FundamentalLimitsCaching} and successfully exploited in~\cite{Wan2020IndexCodingApproach}. Briefly, we recall that an index coding problem~\cite{BarYossef2011IndexCodingSide} consists of a server wishing to deliver $N'$ independent messages to $K'$ users via a basic bottleneck link. Each user $k \in [K']$ has its own \emph{desired message set} $\mathcal{M}_{k} \subseteq [N']$ and has knowledge of its own \emph{side information set} $\mathcal{A}_{k} \subseteq [N']$. The index coding problem can be described by the \emph{side information graph}, a directed graph where each vertex is a message $M_i$ for $i \in [N']$ and where there is an edge from $M_i$ to $M_j$ if $M_i$ is in the side information set of the user requesting $M_j$. The derivation of our converse will make use of the following result from~\cite[Corollary 1]{Arbabjolfaei2013CapacityRegionIndex}. 
\begin{lemma}[{\cite[Corollary 1]{Arbabjolfaei2013CapacityRegionIndex}}]\label{thm: Acyclic Subgraph Converse Bound}
  In an index coding problem with $N'$ messages $M_{i}$ for $i \in [N']$, the minimum number of transmitted bits $\rho$ is lower bounded as
  \begin{equation}
   \rho \geq \sum_{i \in \mathcal{J}} |M_{i}| 
  \end{equation}
for any acyclic subgraph $\mathcal{J}$ of the side information graph.
\end{lemma}

We provide the proof for the non-trivial range\footnote{When $\alpha = 1$ the proof is trivial, since for such case we have only two integer points corresponding to $t \in \{0, 1\}$: when $t = 0$ the load is equal to $K$, and when $t = 1$ each user has enough memory to cache entirely its own FDS and the load is equal to $0$. The case $\alpha = K$ corresponds to the standard MAN scenario already considered in~\cite{Wan2020IndexCodingApproach}.} $\alpha \in [2 : K - 1]$ and for the range $t \in \left[0 : \alpha \right]$. Indeed, for $t = \alpha$ we have $M = \alpha N/K = F\binom{K - 1}{\alpha - 1}$, hence the point $(M, R) = \left( \alpha N/K, 0 \right)$ is trivially achievable as a consequence of each user being able to store the entirety of its FDS.

Since each file is of interest to $\alpha$ users, the first step toward the converse consists of splitting each file in a generic manner into a maximum of $2^{\alpha}$ disjoint subfiles as
\begin{equation}
  W_{f, \mathcal{S}} = \{W_{f, \mathcal{S}, \mathcal{T}} : \mathcal{T} \subseteq \mathcal{S}\}
\end{equation}
for each $\mathcal{S} \subseteq [K]$ with $|\mathcal{S}| = \alpha$ and for each $f \in [F]$, where $W_{f, \mathcal{S}, \mathcal{T}}$ is the subfile of $W_{f, \mathcal{S}}$ cached exactly and only by users in $\mathcal{T}$. Such generic splitting satisfies the uncoded and selfish placement constraint as defined in~\Cref{def: Uncoded and Selfish Cache Placement}.

\subsubsection{Constructing the Index Coding Bound}

Assuming that each user requests a distinct file, we can now identify the index coding problem with $K' = K$ users and $N' = K2^{\alpha - 1} $ messages, such that for any tuple $(\bm{d}, \bm{f})$ the desired message set and the side information set are respectively given by $\mathcal{M}_{k} = \{W_{f_{k}, \mathcal{D}_{k}, \mathcal{T}} : \mathcal{T} \subseteq \mathcal{D}_{k}, k \notin \mathcal{T} \}$ and $\mathcal{A}_{k} = \{W_{f, \mathcal{S}, \mathcal{T}} : f \in [F], \mathcal{S} \subseteq [K], |\mathcal{S}| = \alpha, \mathcal{T} \subseteq \mathcal{S}, k \in \mathcal{T}\}$ for each user $k \in [K]$. To apply \Cref{thm: Acyclic Subgraph Converse Bound}, we are interested in acyclic sets of vertices $\mathcal{J}$ in the side information graph of the problem. In the spirit of~\cite{Wan2020IndexCodingApproach}, we know that the set
\begin{equation}\label{eqn: Acyclic Set of Vertices}
  \bigcup_{k \in [K]} \bigcup_{\mathcal{T} \subseteq [K] \setminus \{u_{1}, \dots, u_{k}\} \cap \mathcal{D}_{u_{k}}} \left\{W_{f_{u_{k}}, \mathcal{D}_{u_{k}}, \mathcal{T}}\right\}
\end{equation}
does not contain any directed cycle for any demand $(\bm{d}, \bm{f})$ and any vector $\bm{u}$, where $\bm{u} = \bm{\pi}^K$ for some $\pi \in S_K$. Consequently, applying \Cref{thm: Acyclic Subgraph Converse Bound} yields the following lower bound
\begin{equation}\label{eqn: Index Coding Lower Bound}
  BR^{\star}_{\textnormal{u}, \textnormal{s}} \geq R(\bm{d}, \bm{f}, \bm{u})
\end{equation}
where $R(\bm{d}, \bm{f}, \bm{u})$ is defined as
\begin{equation}
    R(\bm{d}, \bm{f}, \bm{u}) \coloneqq \sum_{k \in [K]} \sum_{\mathcal{T} \subseteq [K] \setminus \{u_{1}, \dots, u_{k}\} \cap \mathcal{D}_{u_{k}}} \left| W_{f_{u_{k}}, \mathcal{D}_{u_{k}}, \mathcal{T}} \right|.
\end{equation}

\subsubsection{Constructing the Optimization Problem}

Our goal is to create several bounds as the one in~\eqref{eqn: Index Coding Lower Bound} and eventually average all of them to obtain a useful lower bound on the optimal worst-case load. Differently from~\cite{Wan2020IndexCodingApproach}, we will not create a bound for all possible demand vectors and permutations of the $K$ users, but rather we aim to create a bound for a set $\mathcal{C}$ of properly selected demands and for a set $\mathcal{U}_{(\bm{d}, \bm{f})}$ of properly selected permutations for each demand $(\bm{d}, \bm{f}) \in \mathcal{C}$. Hence, we aim to simplify the expression given by
\begin{equation}\label{eqn: Complete Lower Bound}
    \sum_{(\bm{d}, \bm{f}) \in \mathcal{C}} \sum_{\bm{u} \in \mathcal{U}_{(\bm{d}, \bm{f})}} BR^{\star}_{\textnormal{u}, \textnormal{s}} \geq 
    \sum_{(\bm{d}, \bm{f}) \in \mathcal{C}}  \sum_{\bm{u} \in \mathcal{U}_{(\bm{d}, \bm{f})}} R(\bm{d}, \bm{f}, \bm{u}).
\end{equation}

For a permutation $\pi \in H_K$, we consider the demands $(\bm{d}, \bm{f})$ where $\mathcal{D}_k = \{k, \pi((\pi^{-1}(k) + 1) \bmod K), \dots, \pi((\pi^{-1}(k) + \alpha - 1) \bmod K)\}$ and $f_k \in [F]$ for each $k \in [K]$. There is a total of $F^K$ such demands. Considering that the order of $H_K$ is $(K - 1)!$ and that we take $F^K$ demands for each $\pi \in H_K$, we consider $(K - 1)!F^K$ distinct\footnote{Letting $\pi \in H_K$ be a circular permutation ensures that no $\bm{d}$ vector is repeated for any specific $\bm{f}$ vector.} demands in total, denoting by $\mathcal{C}$ the set of such demands. Since the vector $\bm{d}$ depends on some circular permutation $\pi \in H_K$, we will identify from now on each demand with $(\bm{d}_\pi, \bm{f}) \in \mathcal{C}$ to highlight such dependency. For a demand $(\bm{d}_\pi, \bm{f}) \in \mathcal{C}$, we let $\mathcal{U}_{(\bm{d}_\pi, \bm{f})}$ be the set containing the $K$ circular shifts of $\bm{\pi}^K$.

Toward simplifying the expression in~\eqref{eqn: Complete Lower Bound}, we count how many times each subfile $W_{f, \mathcal{S}, \mathcal{T}}$ --- for any $f \in [F]$, $\mathcal{S} \subseteq [K]$ with $|\mathcal{S}| = \alpha$, $\mathcal{T} \subseteq \mathcal{S}$ and $|\mathcal{T}| = t'$ with $t' \in [0 : \alpha]$ --- appears in \eqref{eqn: Complete Lower Bound}. To this end, we need the following lemma.

\begin{lemma}\label{lem: Circular Shift Lemma}
  Let $\pi \in S_K$ be a permutation of the set $[K]$. Consider $k_1, k_2 \in [K]$ such that $k_1 \neq k_2$. Consider
  \begin{equation}
      \ell = (\pi^{-1}(k_2) - \pi^{-1}(k_1)) \bmod K.
  \end{equation}
  Then, out of the $K$ circular shifts of $\bm{\pi}^K$, there is a total of $(K - \ell)$ of them such that $k_1$ appears before $k_2$.
\end{lemma}
\begin{proof}
    Due to lack of space, the proof is relegated to the longer version of this work~\cite{Brunero2021UnselfishCodedCaching}.
\end{proof}

The counting argument proceeds as follows. First, we focus on some subfile $W_{f, \mathcal{S}, \mathcal{T}}$ and we assume that the file $W_{f, \mathcal{S}}$ is requested by some user $k \in \mathcal{S} \setminus \mathcal{T}$. Next, we count the number of demands $(\bm{d}_\pi, \bm{f}) \in \mathcal{C}$ for which $k$ and right-most element included from $\mathcal{T}$ are at distance $\ell$ in $\bm{\pi}^K$. Then, we evaluate how many times the subfile appears in the index coding bounds associated to such demands. After that, noticing that $\ell \in [t' : \alpha  - 1]$, we repeat the procedure for each value of $\ell$. Finally, since the same procedure can be repeated for each $k \in \mathcal{S} \setminus \mathcal{T}$, we multiply the end result by $|\mathcal{S} \setminus \mathcal{T}| = (\alpha - t')$.

Let us focus on the subfile $W_{f, \mathcal{S}, \mathcal{T}}$ for some $f \in [F]$, $\mathcal{S} \subseteq [K]$ with $|\mathcal{S}| = \alpha$, $\mathcal{T} \subseteq \mathcal{S}$ and $|\mathcal{T}| = t'$ for some $t' \in [0 : \alpha]$. For some user $k \in \mathcal{S} \setminus \mathcal{T}$ and for $\ell \in [t' : \alpha - 1]$, there is a total of $a_\ell \coloneqq t'!(\alpha - 1 - t')!(K - \alpha)!\binom{\ell - 1}{t' - 1}F^{K - 1}$ demands $(\bm{d}_\pi, \bm{f}) \in \mathcal{C}$ such that the file $W_{f, \mathcal{S}}$ is requested by such user $k \in \mathcal{S} \setminus \mathcal{T}$ and there are exactly $\ell$ elements in $\bm{\pi}^K$ between $k$ and the right-most element included from $\mathcal{T}$. Let $\mathcal{C}_{k, \ell}$ be the set of such demands. Considering how the acyclic set of vertices in~\eqref{eqn: Acyclic Set of Vertices} is built, the subfile $W_{f, \mathcal{S}, \mathcal{T}}$ appears in the index coding bound induced by each $(\bm{d}_\pi, \bm{f}) \in \mathcal{C}_{k, \ell}$ whenever \emph{all} the elements in $\mathcal{T}$ appear after $k$ in $\bm{u} \in \mathcal{U}_{(\bm{d}_\pi, \bm{f})}$. Since for each demand $(\bm{d}_\pi, \bm{f}) \in \mathcal{C}_{k, \ell}$ there are exactly $\ell$ elements in $\bm{\pi}^K$ separating $k$ and the right-most element from $\mathcal{T}$, we know from~\Cref{lem: Circular Shift Lemma} that there are $(K - \ell)$ vectors $\bm{u} \in \mathcal{U}_{(\bm{d}_\pi, \bm{f})}$ where all the elements in $\mathcal{T}$ appear after $k$. Observing that $\ell \in [t' : \alpha - 1]$ and that such reasoning applies whenever the file $W_{f, \mathcal{S}}$ is requested by any of the $(\alpha - t')$ users in $\mathcal{S} \setminus \mathcal{T}$, the specific subfile $W_{f, \mathcal{S}, \mathcal{T}}$ appears $(\alpha - t')\sum_{\ell = t'}^{\alpha - 1}a_\ell (K - \ell)$ times in~\eqref{eqn: Complete Lower Bound}. The same reasoning applies to any other subfile, hence the expression in~\eqref{eqn: Complete Lower Bound} can be rewritten as
\begin{align}
    R^{\star}_{\textnormal{u}, \textnormal{s}} & \geq \frac{1}{BK!F^K} \sum_{(\bm{d}_\pi, \bm{f}) \in \mathcal{C}}  \sum_{\bm{u} \in \mathcal{U}_{(\bm{d}_\pi, \bm{f})}} R(\bm{d}_\pi, \bm{f}, \bm{u}) \\
                                   & = \sum_{t' = 0}^{\alpha}f(t')x_{t'}
\end{align}
where $f(t')$ and $x_{t'}$ are defined as
\begin{align}
    f(t') & \coloneqq N\frac{(\alpha - t')}{F^K K!}\sum_{\ell = t'}^{\alpha - 1} a_\ell (K - \ell) \\
    0 \leq x_{t'} & \coloneqq \sum_{\mathcal{S} \subseteq [K] : |\mathcal{S}| = \alpha} \sum_{\mathcal{T} \subseteq \mathcal{S} : |\mathcal{T}| = t'} \sum_{f \in [F]} \frac{\left| W_{f, \mathcal{S}, \mathcal{T}} \right|}{NB}.
\end{align}

At this point, we seek to lower bound the minimum worst-case load $R^{\star}_{\textnormal{u}, \textnormal{s}}$ by lower bounding the solution to the following optimization problem
\begin{subequations}\label{eqn: Optimization Problem}
  \begin{alignat}{2}
    &\min_{\bm{x}}  & \quad &\sum_{t' = 0}^{\alpha} f(t') x_{t'}\\
    &\textnormal{subject to}  &  & \sum_{t' = 0}^{\alpha}x_{t'} = 1 \label{eqn: File-Size Constraint}\\
    & & & \sum_{t' = 0}^{\alpha}t'x_{t'} \leq \frac{KM}{N} \label{eqn: Memory-Size Constraint}
  \end{alignat}
\end{subequations}
where \eqref{eqn: File-Size Constraint} and \eqref{eqn: Memory-Size Constraint} correspond to the file-size constraint and the cumulative cache-size constraint, respectively.

\subsubsection{Bounding the Solution to the Optimization Problem}

The function $f(t')$ can be simplified as
\begin{equation}
    f(t') = \frac{\binom{\alpha}{t' + 1} + (K - \alpha)\binom{\alpha - 1}{t'}}{\binom{\alpha}{t'}}
\end{equation}
by means of the well-known hockey-stick identity and after some algebraic manipulations. We encourage the reader to refer to~\cite{Brunero2021UnselfishCodedCaching} for the technical passages. Then, $x_{t'}$ can be considered as a probability mass function and so the optimization problem in~\eqref{eqn: Optimization Problem} can be seen as the minimization of $\mathbb{E}[f(t')]$. We know from~\cite[Lemma 3]{Brunero2021UnselfishCodedCaching} that $f(t')$ is convex and strictly decreasing for increasing $t'$, hence the sequence of inequalities $\mathbb{E}[f(t')] \geq f(\mathbb{E}[t']) \geq f(KM/N)$ holds by Jensen's inequality and the memory-size constraint in~\eqref{eqn: Memory-Size Constraint}. Finally, for $t \coloneqq KM/N$ the converse bound is a piecewise linear curve with the corner points in~\eqref{eqn: Information-Theoretic Lower Bound}, concluding the proof. \qed

\subsection{A Detailed Example for the Converse Bound}\label{sec: A Detailed Example for the Converse Bound}

We present here in detail an example that can help the reader better understand the construction of the bound.

Consider the symmetric $(4, 3, 1)$ FDS structure which involves a file library $\mathcal{L} = \{W_{1, \mathcal{S}} : \mathcal{S} \subseteq [4], |\mathcal{S}| = 3 \}$ consisting of $C = \binom{K}{\alpha} = \binom{4}{3} = 4$ classes of files. Since there is only $F = 1$ file per class, there is a total of $N = FC = 4$ files, hence in this example we make no distinction between files and classes of files. For simplicity, we will here refer to file $W_{1, \mathcal{S}}$ directly as $W_{\mathcal{S}}$, which means that each file is entirely described by a $3$-tuple $\mathcal{S} \subseteq [4]$, and each demand instance is entirely defined by the $\bm{d} = (\mathcal{D}_1, \dots, \mathcal{D}_4)$ vector only. The FDS of each user $k \in [4]$ is given by $\mathcal{F}_{k} = \{W_{S} : \mathcal{S} \subseteq [4], |\mathcal{S}| = 3, k \in \mathcal{S}\}$ and it consists of $F\binom{K - 1}{\alpha - 1} = 3$ files. This example considers $t \in [0 : 3]$, simply because having $M = tN/K = 3$ implies that the point $(M, R) = (3, 0)$ is trivially achievable being each user able to preemptively cache the entirety of its FDS.

Assuming the most general uncoded and selfish cache placement, each file $W_{\mathcal{S}}$ is split into a total of  $2^{\alpha} = 8$ disjoint subfiles as
\begin{equation}
    W_{\mathcal{S}} = \{W_{\mathcal{S}, \mathcal{T}} : \mathcal{T} \subseteq \mathcal{S}\}, \quad \forall \mathcal{S} \subseteq [4] : |\mathcal{S}| = 3
\end{equation}
where again $W_{\mathcal{S}, \mathcal{T}}$ is the subfile of $W_\mathcal{S}$ cached by users in $\mathcal{T}$.

\subsubsection{Constructing the Index Coding Bound}

For any given demand $\bm{d} = (\mathcal{D}_1, \dots, \mathcal{D}_4)$ where the $k$-th user asks for a distinct file $W_{\mathcal{D}_{k}}$ for each $k \in [4]$, we consider the index coding problem with $K' = K = 4$ users and $N' = K2^{\alpha - 1} = 16$ messages, where each user $k \in [4]$ has a desired message set $\mathcal{M}_{k} = \{W_{\mathcal{D}_{k}, \mathcal{T}} : \mathcal{T} \subseteq \mathcal{D}_{k}, k \notin \mathcal{T} \}$ and a side information set $\mathcal{A}_{k} = \{W_{\mathcal{S}, \mathcal{T}} : \mathcal{S} \subseteq [4], |\mathcal{S}| = 3, \mathcal{T} \subseteq \mathcal{S}, k \in \mathcal{T}\}$.

If we identify again in the side information graph of the problem the acyclic set of vertices as in~\eqref{eqn: Acyclic Set of Vertices}, applying \Cref{thm: Acyclic Subgraph Converse Bound} yields the following lower bound
\begin{equation}\label{eqn: Index Coding Lower Bound 2}
  BR^{\star}_{\textnormal{u}, \textnormal{s}} \geq R(\bm{d}, \bm{u})
\end{equation}
where now we define
\begin{equation}
    R(\bm{d}, \bm{u}) \coloneqq \sum_{k \in [4]} \sum_{\mathcal{T} \subseteq [4] \setminus \{u_{1}, \dots, u_{k}\} \cap \mathcal{D}_{u_{k}}} \left| W_{\mathcal{D}_{u_{k}}, \mathcal{T}} \right|.
\end{equation}
Recall that~\eqref{eqn: Index Coding Lower Bound 2} holds for any demand $\bm{d}$ and any $\bm{u} = \bm{\pi}^4$ for some $\pi \in S_4$.

\subsubsection{Constructing the Optimization Problem and Bounding its Solution}

As described in~\Cref{sec: Main Proof}, we let $\mathcal{C}$ be the set of distinct demands where we have a $\bm{d}_\pi$ vector for each circular permutation $\pi \in H_4$. The number of such permutations is $(K - 1)! = 6$. For simplicity, we denote by $\pi_i$ the $i$-th circular permutation for $i \in [6]$, where such circular permutations are given by
\begin{alignat}{2}
    \bm{\pi}_1^4 & = (1, 2, 3, 4) & \quad \bm{\pi}_4^4 & = (1, 3, 4, 2) \\
    \bm{\pi}_2^4 & = (1, 2, 4, 3) & \quad \bm{\pi}_5^4 & = (1, 4, 2, 3) \\
    \bm{\pi}_3^4 & = (1, 3, 2, 4) & \quad \bm{\pi}_6^4 & = (1, 4, 3, 2).
\end{alignat}
None of the above permutations can be obtained as a rotation of any of the others, hence this ensures to have in $\mathcal{C}$ the following distinct demand vectors
\begin{alignat}{2}
    \bm{d}_{\pi_1} & = (123, 234, 134, 124) & \ \bm{d}_{\pi_4} & = (134, 123, 234, 124) \\
    \bm{d}_{\pi_2} & = (124, 234, 123, 134) & \ \bm{d}_{\pi_5} & = (124, 123, 134, 234) \\
    \bm{d}_{\pi_3} & = (123, 124, 234, 134) & \ \bm{d}_{\pi_6} & = (134, 124, 123, 234)
\end{alignat}
with $\mathcal{U}_{\bm{d}_{\pi_i}}$ containing the $4$ circular shifts of $\bm{\pi}^4_i$. For instance, it is $\mathcal{U}_{\bm{d}_{\pi_1}} = \{(1, 2, 3, 4), (2, 3, 4, 1), (3, 4, 1, 2), (4, 1, 2, 3)\}$. 

Now, we create a bound as in~\eqref{eqn: Index Coding Lower Bound 2} for each $\bm{d}_{\pi_i}$ with $i \in [6]$ and for each $\bm{u} \in \mathcal{U}_{\bm{d}_{\pi_i}}$ for a given $\bm{d}_{\pi_i}$, we sum all such bounds and we obtain the expression given by
\begin{equation}\label{eqn: Complete Lower Bound 2}
    \sum_{i \in [6]} \sum_{\bm{u} \in \mathcal{U}_{\bm{d}_{\pi_i}}} BR^{\star}_{\textnormal{u}, \textnormal{s}} \geq 
    \sum_{i \in [6]} \sum_{\bm{u} \in \mathcal{U}_{\bm{d}_{\pi_i}}} R(\bm{d}_{\pi_i}, \bm{u}).
\end{equation}
We simplify~\eqref{eqn: Complete Lower Bound 2} by counting how many times each subfile $W_{\mathcal{S}, \mathcal{T}}$ --- for any $\mathcal{S} \subseteq [4]$ with $|\mathcal{S}| = 3$, $\mathcal{T} \subseteq \mathcal{S}$ and $|\mathcal{T}| = t'$ with $t' \in [0 : 3]$ --- appears in~\eqref{eqn: Complete Lower Bound 2}.

As an example, we focus on the subfile $W_{123, 2}$. First of all, we notice that such subfile may appear to the RHS of~\eqref{eqn: Index Coding Lower Bound 2} whenever $W_{123}$ is requested by any user $k \in \mathcal{S} \setminus \mathcal{T} = \{1, 3\}$. Assume that $W_{123}$ is requested by user~$1$, which means we consider the bounds with demands $\bm{d}_{\pi_1}$ and $\bm{d}_{\pi_3}$. Denoting by $\ell$ the distance\footnote{We remind that such distance represents the number of elements which separate index $1$ and index $2$, including the latter.} between user~$1$ and user~$2$ in $\bm{\pi}^4_i$ for $i \in \{1, 3\}$, we notice that $\ell \in \{1, 2\}$. We see that there is only $\bm{\pi}^4_1$ for $\ell = 1$, hence $W_{123, 2}$ appears $(K - \ell) = 3$ times in the bound with demand $\bm{d}_{\pi_1}$ and permutations in $\mathcal{U}_{\bm{d}_{\pi_1}}$. Similarly, there is only $\bm{\pi}^4_3$ for $\ell = 2$, hence $W_{123, 2}$ appears $(K - \ell) = 2$ times in the bound with demand $\bm{d}_{\pi_3}$ and permutations in $\mathcal{U}_{\bm{d}_{\pi_3}}$. Thus, summing over the possible values of $\ell$, the subfile $W_{123, 2}$ appears $3 + 2 = 5$ times in the bounds with demands $\bm{d}_{\pi_1}$ and $\bm{d}_{\pi_3}$, and their relative permutations of users. The same rationale follows for the bounds built with $\bm{d}_{\pi_2}$, $\bm{d}_{\pi_6}$ and their relative permutations, i.e., the bounds where file $W_{123}$ is requested by user~$3$. Hence, the subfile $W_{123, 2}$ appears in~\eqref{eqn: Complete Lower Bound 2} a total of $2 \times (3 + 2) = 10$ times. The same holds for any other subfile cached at only one user and a similar counting argument can be made for arbitrary $|\mathcal{T}| = t'$ with $t' \in [0 : 3]$, as described in~\Cref{sec: Main Proof}. Thus, we can formulate an optimization problem as in~\eqref{eqn: Optimization Problem}, bounding its solution by means of Jensen's inequality and convexity of $f(t')$. The resulting bound is a piecewise linear curve with corner points $(M, R_{\textnormal{LB}}) = \left(t, (3 - t)/(1 + t) + 1 - t/3 \right)$ for each $t \in [0 : 3]$.

\section{Conclusions}\label{sec: Conclusions}

In this work, we investigated the effects that selfish caching can have on the optimal worst-case communication load in the coded caching framework. We proposed a general FDS structure that seeks to capture the degree of intersection between the interests of the different users and for such generic structure we provided a novel information-theoretic converse on the minimum worst-case communication load under uncoded and selfish cache placement. Such bound definitively resolves the question of whether selfish caching is beneficial or not in general. Indeed, for the proposed symmetric FDS structure, the presented converse not only reveals that any non-zero load brought about by symmetrically selfish caching is always (with the exception of the extreme points of $t$) strictly worse than the optimal load guaranteed in the unselfish scenario, but also offers a definitive proof that selfish coded caching can yield unbounded losses over non-selfish coded caching, as it is proved in the longer version of this work~\cite{Brunero2021UnselfishCodedCaching}.

\bibliographystyle{IEEEtran}
\bibliography{references}

\end{document}